\documentclass[structabstract]{aa} 
\usepackage{graphicx}
\usepackage{txfonts}
\begin{document}
 \title{HD~51106 and HD~50747: an ellipsoidal binary and a triple system observed with CoRoT\thanks{Based on observations obtained by CoRoT, a space project operated by the French Space Agency (CNES) with the participation of the Science Program of ESA, ESTEC/RSSD, Austria, Belgium, Brazil, Germany, and Spain. Also based on observations made with ESO telescopes at the La Silla Observatory under the ESO large program LP178.D-0361, at the Observatoire de Haute Provence, INSU/CNRS, France, 
and on observations collected at the Centro Astron\'omico Hispano Alem\'an (CAHA) at Calar Alto, operated jointly by the Max-Planck Institut f\"ur Astronomie and the Instituto de Astrof\'{\i}sica de Andaluc\'{\i}a}.}
 \subtitle{}

 \author{
 N. Dolez
 \inst{1}
 \and
 S. Vauclair
 \inst{1}
 \and
 E. Michel
 \inst{2}
 \and
 A. Hui-Bon-Hoa
 \inst{1}
 \and
 G. Vauclair
 \inst{1}
 \and
 D. Le Contel
 \inst{3}
 \and
 P. Mathias
 \inst{3}
 \and
 E. Poretti
 \inst{4}
 \and
 P.J. Amado
 \inst{6}
 \and
 M. Rainer
 \inst{4}
 \and
 R. Samadi
 \inst{2}
 \and
 A. Baglin
 \inst{2}
 \and
 C. Catala
 \inst{2}
 \and
 M. Auvergne
 \inst{2}
 \and
 K. Uytterhoeven
 \inst{4,5}
 \and
 J.-C. Valtier
 \inst{3}
 }

 \institute{Laboratoire d'Astrophysique de Toulouse-Tarbes, Universit\'e de Toulouse, CNRS, 14 avenue Edouard Belin, F31400
Toulouse, France
 \and
 Observatoire de Paris, LESIA, FRE 2461, ­F92195, Meudon, France
 \and
 Dpt Fizeau, UMR 6525 Observatoire de la C\^ote d'Azur/CNRS, BP 4229, F06304 Nice Cedex 4, France
 \and
 INAF ­Osservatorio Astronomico di Brera, Via Bianchi 46, I­23807 Merate, Italy
 \and
 Laboratoire AIM, CEA/DSM CNRS Universit\'e Paris Diderot; CEA, IRFU, SAp, centre de Saclay, F91191, Gif-sur-Yvette, France
 \and
 Instituto de Astrof\'{\i}sica de Andaluc\'{\i}a-CSIC,  P.O. Box 3004. Granada E-18080, Spain
}

 \date{Received; accepted }

 
 \abstract
 {We present an analysis of the observations of HD~51106 and HD~50747 by the satellite CoRoT, obtained during its initial run, and of the spectroscopic preparatory observations.}
 {We complete an analysis of the light curve, extract the main frequencies observed, and discuss some preliminary interpretations about the stars.}
 {We used standard Fourier transform and pre-­whitening methods to extract information about the periodicities of the stars.}
 {HD 51106 is an ellipsoidal binary, the light curve of which can be completely explained by the tidal deformation of the star and smaller secondary effects. 
HD 50747 is a triple system containing a variable star, which exhibits many modes of oscillation with periods in the range of a few hours. On the basis of this period range and the analysis of the physical parameters of the star, we conclude that HD~50747 is a $\gamma$-Doradus star.}
 {}


 \maketitle
%

\section{Introduction}

 Both HD~51106 and HD~50747 have been observed with CoRoT (Baglin et al. 2006). They were secondary targets observed in the asteroseismology field during the initial run of the satellite.

The choice of HD~51106 was based on its spectral type (Am) in our search for possible variability. Since very few Am stars oscillate (as reviewed in  Kurtz, 2000), it is thus very interesting to investigate more thoroughly the link between pulsations and chemical pecularities with both the high sensitivity and frequency resolution of CoRoT.
HD~51106 has a magnitude of 7.35

HD~50747 was the brightest secondary target of this run, and has a spectral type A4IV and a visual magnitude of 5.45. It was included in this study because of a previous classification as an Am star. Although its spectral classification has not yet been completely settled, this was a promising object for searching for variability.

In preparation for the CoRoT mission, extensive high resolution spectroscopy was completed to obtain more precise physical parameters of these stars. 

Section 2 presents an analysis of the pre-­CoRoT spectroscopic observations of the stars, i.e., orbital parameters, fundamental parameters, abundance analysis. In Sect. 3, we describe the CoRoT data, the method of analysis, and some characteristics of the Fourier transform. A discussion of the significance of the observations, in terms of variable stellar class for HD~50747, and binarity effects for HD~51106, is given in Sect. 4. 

\section{Ground-based spectroscopy}
\subsection{The data}
The spectra that we used were of three kinds, all obtained in the framework of the CoRoT seismology ground-based observation working group. The logbook of the observations is provided in Table~\ref{table:1}, which provides the names of the instruments used, the date of the beginning and end of observations, the number of spectra obtained and their mean S/N, and the initials of the observers. The SOPHIE spectrograph is mounted on the 1.93 m telescope at Observatoire de Haute-Provence, France, FEROS on the 2.2-m ESO/MPI telescope at ESO, Chile (Poretti et al., 2007), and FOCES on the 2.2-m  telescope at the Observatory of Calar Alto in Almeria, Spain. \\

Spectra were reduced using the software of each observing team.
To improve the signal-to-noise ratio,
we computed the correlation profiles using the least square deconvolution (LSD)
method (Donati et al. 1997). The spectral line list was chosen
from the VALD database (Piskunov et al. 1995) concerning a template (Teff,logg)=(8000,4). 
Common for both stars, this template matches the stellar parameters as closely as possible. 
All stellar lines apart from those of H and He are included in the deconvolution.
The correlation profiles derived have a signal-to-noise ratio above 700. 
For both stars, they are presented in Figs. 1 \& 2.

\begin{table}[ht]
\caption{Logbook of the observations. The {\it JD} are given in julian days (-2454000).
} 
\label{table:1} 
\centering 
\begin{tabular}{l l l l} 
\hline\hline 
$Instrument$ &$SOPHIE$     &$FEROS$      &$FOCES$      \\
  $Resolution$  &70000 &48000 &35000 \\
$JD_{start}$ &112.382  &103.657  &  76.618  \\
$JD_{end}$ & 132.336 &128.701 &  78.610 \\
$N_{HD50747}$ & 14    & 18    &  7 \\
$N_{HD51106}$ & 13 & 13 & 4 \\
 $S/N$ & 60 & 80 & 60 \\
$Observers$ & $PM, JCV$ & $MR, KU$ & $PJA$\\
\hline 
\end{tabular}
\end{table}

\subsection{Orbital parameters}

To determine the orbital parameters, we measured the velocities of the different
components using  a multi-Gaussian fit performed on the corrrelation profiles.
The solution for the binary orbits was then computed using a Lehmann-Filhes' method-based code that in addition 
calculated simultaneously the internal errors  (Lehman-Filhes 1894). 
Concerning HD~51106, it appears that the eccentricity
is very low
($e < 0.01$), so we imposed a circular orbit. Although low, the eccentricity
of HD~50747 is
significant, and we retained this value when computing the orbital
ephemeris. Note that we only used the velocities computed from a Gaussian fit.
The binary parameters for both stars are given in Tables 2 \& 3, and the
corresponding orbital solutions
are represented in Figs. 3 \& 4.\\

The case of HD~50747 is complicated because this star exhibits complex line profiles, with two narrow components belonging to a spectral binary (SB2) system, and at least a broad component whose variation is very difficult to establish (Fig.~\ref{figure:2}). This could be interpreted as being a triple system, with two close components and a third, much brighter star responsible for at least 80\% of the total luminosity of the system. The orbital parameters of the two close components are listed in { Table~\ref{table:3}}. The bright star has rapid rotation (see below), but no measurable radial velocity variation: we can therefore infer that this star is far away from the two others and the corresponding orbital motion is too slow to be measured with the current spectroscopic dataset.

 \begin{figure}
\resizebox{\hsize}{!}{\includegraphics{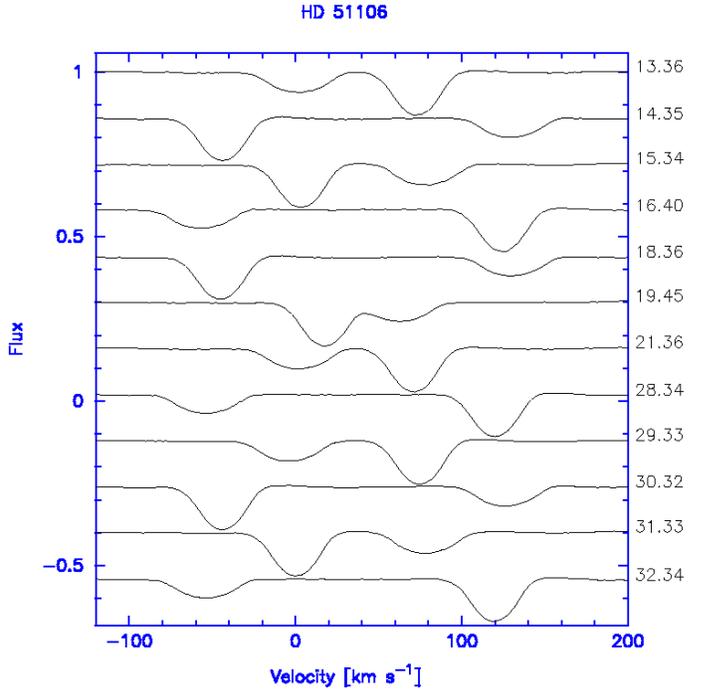}}
 \caption{Correlation profiles of HD~51106, from SOPHIE spectra. The right axis is labelled in julian days (+2454100). The profiles are offset for clarity.}
 \label{figure:1}%
 \end{figure}

\begin{figure}
\resizebox{\hsize}{!}{\includegraphics{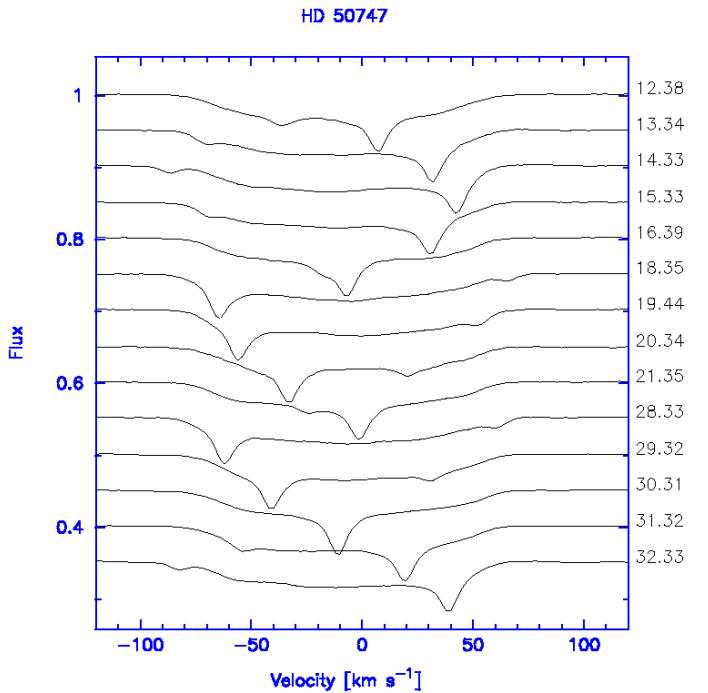}}

 \caption{Correlation profiles of HD~50747, from SOPHIE spectra. The right axis is labelled in julian days (+2454100).}
 \label{figure:2}
 \end{figure}

\begin{table}[ht]
\caption{Parameters of the binary orbit for the star HD~51106.  The mass ratio is $M_1/M_2$ = 1.12.} 
\label{table:2} 
\centering 
\begin{tabular}{r c l} 
\hline\hline 
$P$ &=& 4.006 $\pm$ 0.001 d \\
$T_0$ &=& 2454193.9 $\pm$ 0.7 d \\
$\gamma$ &=& 39.2 $\pm$ 0.3 km~s$^{-1}$\\
$e$ &=& $0$ (assumed)\\
$K_1$ &=& 90.5 $\pm$ 0.7 km~s$^{-1}$\\
$K_2$ &=& 101.2 $\pm$ 0.7 km~s$^{-1}$\\
$\omega_1$ &=& 90.0 $\pm$ 0.0\degr\\
$a_1 \sin i$ &=& (4.98 $\pm$ 0.04) $10^6$ km \\
$a_2 \sin i$ &=& (5.58 $\pm$ 0.04) $10^6$ km \\
$M_1 \sin^3 i$ &=& 1.55 $\pm$ 0.02 M$_\odot$ \\
$M_2 \sin^3 i$ &=& 1.38 $\pm$ 0.02 M$_\odot$ \\
\hline 
\end{tabular}
\end{table}

\begin{table}[ht]
\caption{Parameters of the binary orbit for the star HD~50747.  The mass ratio is $M_1/M_2$ = 1.40.} 
\label{table:3} 
\centering 
\begin{tabular}{r c l} 
\hline\hline 
$P$ &=& 9.25 $\pm$ 0.01 d \\
$T_0$ &=& 2454165.8 $\pm$ 0.9 d \\
$\gamma$ &=& -10.9 $\pm$ 0.2 km~s$^{-1}$\\
$e$ &=& 0.073   $\pm$ 0.004\\
$K_1$ &=& 54.1 $\pm$ 0.3 km~s$^{-1}$\\
$K_2$ &=& 75.5 $\pm$ 0.3 km~s$^{-1}$\\
$\omega_1$ &=& 95.0 $\pm$ 2.3\degr\\
$a_1 \sin i$ &=& (6.85 $\pm$ 0.04) $10^6$ km \\
$a_2 \sin i$ &=& (9.57 $\pm$ 0.04) $10^6$ km \\
$M_1 \sin^3 i$ &=& 1.21 $\pm$ 0.01 M$_\odot$ \\
$M_2 \sin^3 i$ &=& 0.86 $\pm$ 0.01 M$_\odot$ \\
\hline 
\end{tabular}
\end{table}

 \begin{figure}
\resizebox{\hsize}{!}{\includegraphics{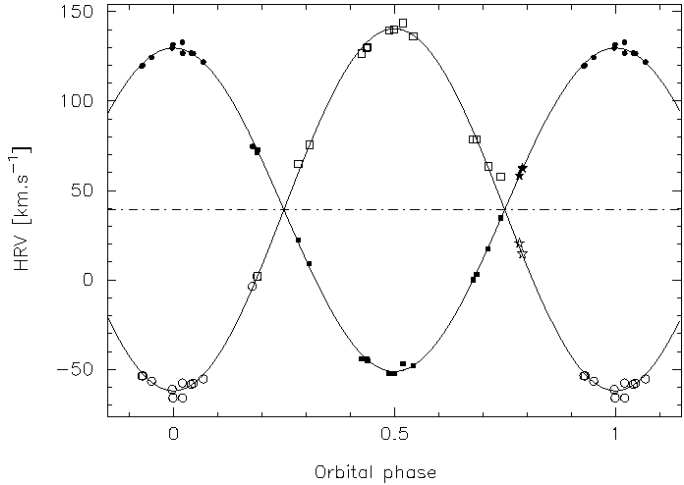}}
 \caption{Orbit of the star HD~51106. The horizontal dashed-line represents the heliocentric velocity of the system. The different symbols indicate the radial velocities obtained with SOPHIE (circles), FEROS (squares), and FOCES (stars) spectrographs.}
 \label{figure:3}%
 \end{figure}

 \begin{figure}
\resizebox{\hsize}{!}{\includegraphics{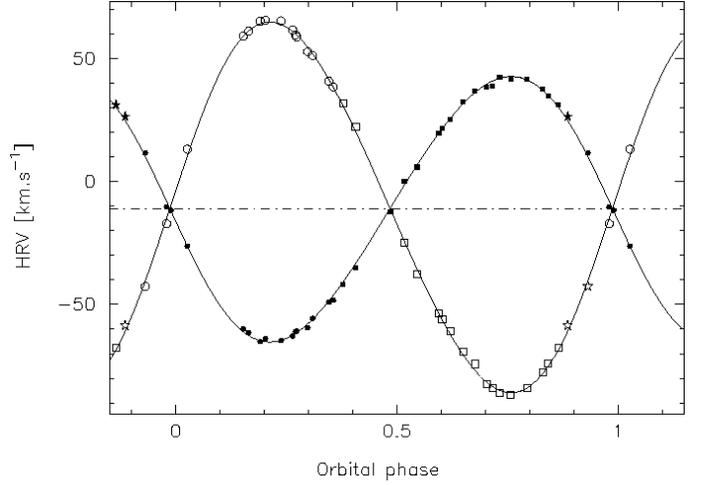}}
 \caption{Orbit of the star HD~50747. The horizontal dashed-line represents the heliocentric velocity of the system. Symbols are the same as those presented in Fig. 3.}
 \label{figure:4}%
 \end{figure}

\subsection{Fundamental parameters}
HD~51106 is a double-lined spectroscopic binary, and the two components have similar masses. We assumed that both components of the system have similar fundamental parameters and estimated them using the composite Str\"omgren photometry.
We used two sets of photometric data: those of Hauck \& Mermilliod (1998), and those of the Gaudi database (Solano et al. 2005). These were dereddened using the code of Moon \& Dworetsky (1985) and Moon (1985) (Table~\ref{table:4}).
 The fundamental parameters for HD~51106 are $T_\mathrm{eff}=7870~K$ and $\log~g=3.94$ for the first set of data, and $T_\mathrm{eff}=8050~K$ and $\log~g=4.05$ for the second set.

HD~50747 is a triple system, consisting of one bright component, and two much fainter components.
To date, we have been unable to disentangle the three spectra and derive the fundamental parameters of the individual components. We can only estimate the fundamental parameters of the brightest component, using the Str\"omgren photometry, because this component contributes about 80\% of the total luminosity. 
By continuing to apply the Moon \& Dworetsky (1985) and Moon (1985) codes and the photometric data of Hauck \& Mermilliod (1998) and of the Gaudi database, the fundamental parameters obtained for HD50747 are  $T_\mathrm{eff}=7980~K$ and $\log~g=3.45$ with the first set of data, and $T_\mathrm{eff} = 7810~K$ and $\log~g=3.32$ with the second set. The discrepancies arise mostly from the difference in H$\beta$ for the effective temperature, and in $c_0$ for the surface gravity.

\begin{table}[ht]
\caption{Dereddened Str\"omgren photometric data. H \& M represents Hauck \& Mermilliod (1998).}
\label{table:4}
\centering
\begin{tabular}{c c c c c}
\hline\hline
\multicolumn{4}{c}{Data for HD51106:} & \\
\hline
  &$(b-y)_0$&$m_0$&$c_0$&H$\beta$\\
 H \& M  &0.119&0.245&0.914&2.820 (estimated)\\
 Gaudi &0.108&0.242&0.922&2.839\\
\hline\hline
\multicolumn{4}{c}{Data for HD50747:} & \\
\hline
  &$(b-y)_0$&$m_0$&$c_0$&H$\beta$\\
 H \& M  &0.084&0.176&1.129&2.833\\
 Gaudi &0.084&0.176&1.135&2.815\\
\hline
\end{tabular}
\end{table}

\subsection{Abundance analysis}
Attempts were made to fit synthetic spectra to the observed data. We used the grids computed with the ATLAS9 (Kurucz 1993) LTE model atmosphere, extracting a 8000~K, $\log~g = 4$ model. The synthesis was performed using a code developed by V. Tsymbal (private communication) in LTE: the spectrum is convolved with an adequate instrumental and rotational profile.

\begin{figure}[ht]
\resizebox{\hsize}{!}{\includegraphics{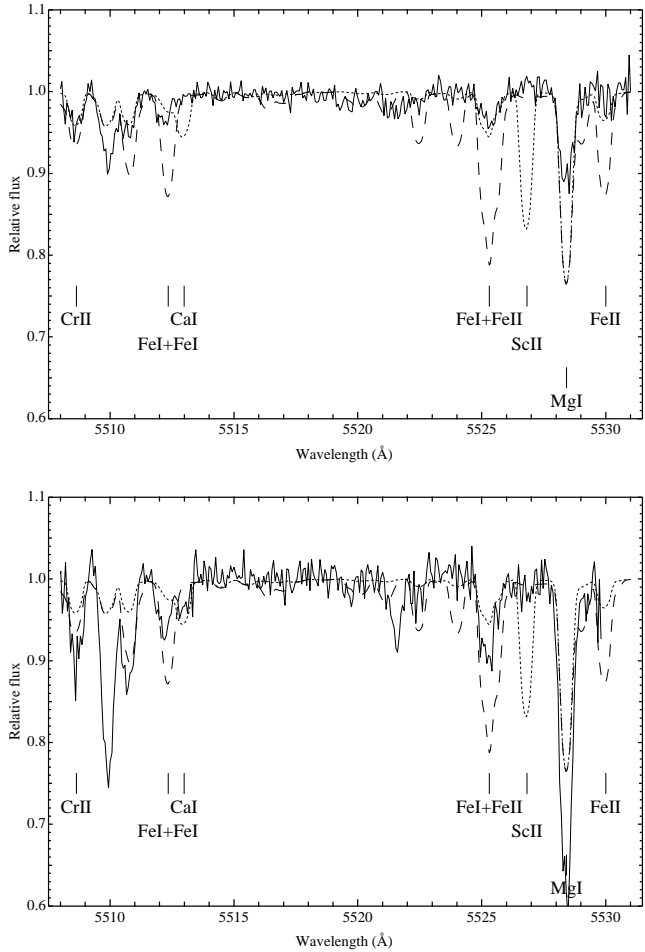}}
 \caption{Observed and synthetic spectra for each of the components 
of HD~51106. The solid line represents the disentangled spectra, whereas 
dotted and dashed-lines show the solar abundance spectra and the Am 
abundance spectra, respectively.}
 \label{figure:5}%
 \end{figure}

For HD~51106, as a first attempt, we chose the spectral interval $\lambda\lambda 5500-5535\AA$, because the program used to disentangle the spectra (CRES, Ilijic 2004) can only deal with a very small span of data. This interval was chosen since it includes lines of several chemical elements of interest in the Am phenomenon, namely Ca, Sc, and Fe. For the disentanglement, we assumed that the luminosity ratio is around $1.45$ according to the standard mass-luminosity relation on the main sequence.
 For each component, we compared the disentangled spectrum with two synthetic spectra: the first one has solar abundances for all the elements, whereas the second has Ca and Sc underabundant compared to solar by a factor of 100, and Fe overabundant by a factor of 10.\\
In Fig.~\ref{figure:5}, we see that in both components, Ca and Sc are clearly underabundant (by a factor of 0.01), and Fe is almost solar or only slightly overabundant. Further investigation is needed to draw more precise conclusions. Nevertheless, confirming what is suggested by spectral classification, this system consists of two Am stars.

As a by-product, using the shape of the lines in the correlation profiles, we estimated the rotational velocity of the two components ($v~\sin~i$ around 20~km~s$^{-1}$).

For HD~50747, we only tried to fit the lines of the brightest component, because, as indicated previously, we could not disentangle the three spectra of the system at this stage. A solar abundance spectrum seems convenient, excluding any strong abundance anomaly.
We could also estimate the rotational velocity of the brightest component ($v~\sin~i$ = 80~km~s$^{-1}$), and those of the fainter components (around 10~km~s$^{-1}$).

\section{CoRoT Observations}
HD~51106 and HD~50747 were observed continuously over 60 days during the 
initial run of CoRoT which began on February 7, 2007. The stars were secondary targets of the CoRoT asteroseismology field. 
\subsection{HD 51106 light curve}
\begin{figure}
\resizebox{\hsize}{!}{\includegraphics{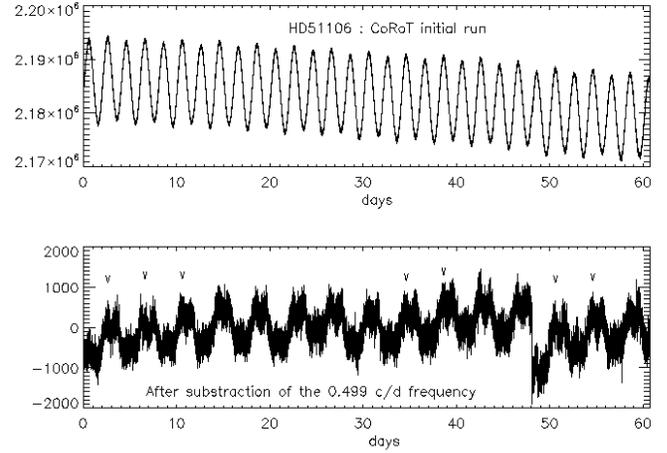}}
 \caption{Upper panel: raw 'N2' light curve of HD~51106.
Lower panel~: residual light curve after polynomial correction of the trend and \hbox{subtraction} of the first frequency (0.499 d$^{-1}$). The ticks show the position of some of the even (e.g., higher) maxima of the light curve, and are here to make the phase difference clearly visible. Note the instrumental jump around t=48 days.}
 \label{figure:6}
\end{figure}

In the CoRoT light curve (Fig.~\ref{figure:6}, upper panel), one immediately notices a very regular oscillation, a slight modulation with twice that period, and a slow temporal drift which can be attributed to CCD ageing (cf Auvergne et al. 2009, this issue). After substracting the drift, using a polynomial fit of the data, we proceeded to extract the frequencies using the Period04 software (Lenz and Breger 2005). 

\begin{figure}
\resizebox{\hsize}{!}{\includegraphics{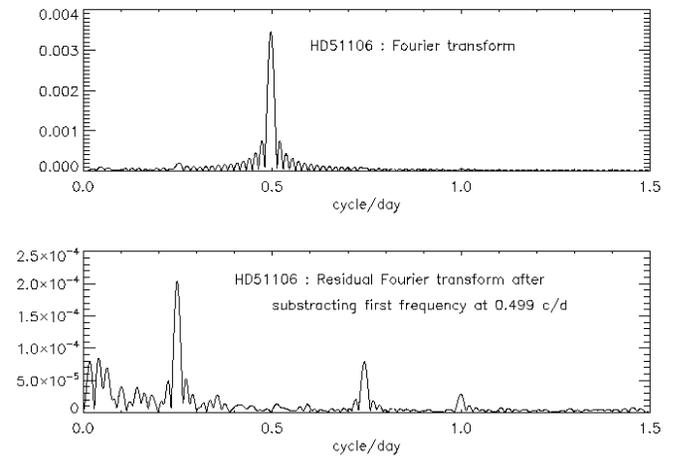}}
 \caption{Upper panel: Fourier transform of the de­trended light curve. Lower panel: residual Fourier transform after extracting the main frequency (0.499 d$^{-1}$). The highest peaks correspond to the orbital period of the binary, and the others are linear  combinations or harmonics. The phase of the oscillations at 0.499 d$^{-1}$, 0.25 d$^{-1}$, and 0.75 d$^{-1}$ are 0.95, 0.51, and 0.52, respectively.}
 \label{figure:7}
 \end{figure}

The Fourier transform detects two frequencies (Fig.~\ref{figure:7}, upper panel) close to 0.499 d$^{-1}$ and 0.25 d$^{-1}$. We analysed the data, using pre­whitening methods (included in the Period04 code). Apart from the above-mentioned frequencies, there are no other peaks in the power spectrum above the noise level, except for peaks that are probably harmonics or linear combinations of those two (Fig.~\ref{figure:7}, lower panel) peaks caused by the spectral window, and low frequency peaks related to the residual long period drift of the data and jumps in the light curve. All significant frequencies, even the most conspicuous at 0.499 d$^{-1}$, can indeed be described as harmonics of the 0.25 d$^{-1}$ one. 

It is interesting to consider the residual light curve after substracting the main frequency (Fig.~\ref{figure:6}, lower panel). One can see that the 0.25 d$^{-1}$ oscillation is almost in phase with the main frequency (its maximum being offset by about 0.15 period). We can also clearly see that the next harmonic (0.75 d$^{-1}$) has a phase exactly opposite to that of the 0.25 d$^{-1}$ oscillation, with maxima of the former falling exactly within minima of the latter. This combination results in a difference between even and odd minima, and also between even and odd maxima of the light curve.\\

There is a very conspicuous jump in the data at around 48 days: this is clearly an instrumental effect that has been only partially corrected during the pre-­reduction stages completed while obtaining the 'N2' data set. This effect appears when a CCD pixel is affected by a cosmic ray. At our level of analysis, we did not attempt to improve this, because it affects only very low frequencies (and somehow the noise level), and does not influence the domain of frequencies of interest for this binary.

\subsection{HD~50747 light curve}
\begin{figure}
\resizebox{\hsize}{!}{\includegraphics{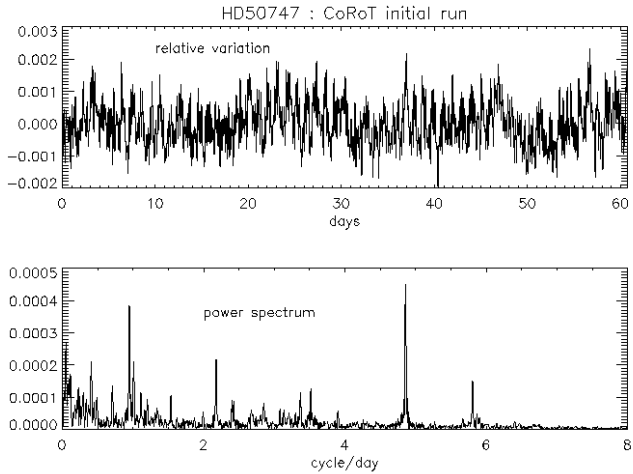}}
 \caption{Upper panel: « detrended » light curve of HD 50747. 
 Lower panel: Power Spectrum (square root).}
 \label{figure:8}%
 \end{figure}

The CoRoT light curve clearly exhibits variations, of an amplitude of about one thousandth of the total luminosity (Fig.~\ref{figure:8}, upper panel). 

The Fourier transform (Fig.~\ref{figure:8}, lower panel) exhibits many frequencies. The two main peaks are at F1=4.865 d$^{-1}$ and F2=0.956 d$^{-1}$ (periods of 4.93 hours and 25.1 hours, respectively). We analysed the data using the Period04 code (Lenz and Breger 2005) and found more than 40 significant frequencies: the frequencies listed in Table~\ref{table:5} all have an amplitude more than 3 times larger than the noise level, which is estimated to be 15.$\mu$mag in the domain 1 d$^{-1}$ - 6 d$^{-1}$ . We note that no significant frequency was found above 6 d$^{-1}$, except for peaks caused by the spectral window of the observations. 

In this table, we did not include some very low frequency peaks, which we believe are caused by instrumental effects, such as, a slow drift of data or hot pixels correction. 
We could not detect any clear effect of the orbital motion of the stars in the dataset: the orbital frequency  (about 0.108 d$^{-1}$) and its first harmonics were lost in the low frequency noise. We also tried to identify a modulation of the oscillations by the orbital motion, which would appear as lateral peaks around, for example, F1 (4.86 d$^{-1}$), separated from it by the orbital frequency. From the parameters of the star system, we calculated that the amplitude of those peaks should be about $0.5\%$ of the amplitude of F1, and this is lower than the noise in this part of the spectra, even after substracting the oscillation F1 itself.

Two frequencies were identified as linear combinations of the two main frequencies (sum and difference: F6 and F26 in Table~\ref{table:5}). The frequency F22 is probably an observation artefact, related to the combination of satellite orbit and Earth rotation. This frequency at 2.008 d$^{-1}$ remarkably appears also in the spectrum of HD~51106. Since it is very close to an harmonic of the orbital period of the star, it could easily be confused with it. The high resolution of CoRoT, because of the 60 days duration of the observation, helped to prevent this confusion.

\begin{table}
\caption{Period04 analysis results of HD~50747.} 
\label{table:5} 
\centering 
{\scriptsize
\begin{tabular}{rrrrr} 
\hline\hline 
& \multicolumn{2}{c}{ Frequency}  & Amplitude & Comment\\
&$\mu$Hz&d$^{-1}$&$\mu$mag&\\ 
\hline 
F1&56.309&4.865&465.326&\\
F2&11.070&0.956&419.315&\\
F3&4.787&0.414&218.525&\\
F4&25.381&2.193&225.250&\\
F5&11.804&1.020&200.361&\\
F6&67.383&5.822&157.676&F1+F2\\
F7&8.262&0.714&137.349&\\
F8&40.985&3.541&155.490&\\
F9&12.997&1.123&115.464&\\
F10&39.320&3.397&117.916&\\
F11&5.710&0.493&111.093&\\
F12&17.909&1.547&105.581&\\
F13&28.056&2.424&86.470&\\
F14&14.019&1.211&98.215&\\
F15&33.271&2.875&95.513&\\
F17&28.271&2.443&84.361&\\
F18&15.628&1.350&57.912&\\
F19&35.905&3.102&60.125&\\
F20&31.232&2.698&66.914&\\
F21&5.230&0.452&77.300&\\
F22&23.246&2.008&60.237&artefact?\\
F23&36.602&3.162&57.721&\\
F24&37.385&3.230&56.063&\\
F25&10.740&0.928&61.108&\\
F26&45.144&3.900&54.500&F1-F2\\
F27&13.658&1.180&62.307&\\
F28&4.309&0.372&59.474&\\
F29&31.437&2.716&49.910&\\
F30&32.759&2.830&56.642&\\
F31&15.786&1.364&49.791&\\
F32&68.125&5.886&48.071&\\
F33&33.575&2.901&48.838\\
F34&4.913&0.424&60.885\\
F35&15.314&1.323&48.604\\
F36&30.782&2.660&47.912\\
F37&13.216&1.142&45.118\\
F38&40.863&3.531&70.136\\
F39&7.422&0.641&40.011\\
F40&32.265&2.788&41.449\\
F41&68.548&5.923&39.074\\
\hline 
\end{tabular}
}
\end{table}

\section{Interpretation}
Our results for the two stars systems are very different, albeit interesting.

\subsection{HD~51106}
The light curve of this object is convincingly characteristic of an « ellipsoidal » binary, with a periodic signal at half the orbital period, plus a smaller signal at the orbital period. 
The CoRoT observation confirms to good accuracy (superior to $10^{-4}$) the orbital period obtained from the ground (Table~\ref{table:2}). The main period of two days is caused by the deformation of the stars into an ellipsoidal shape, caused by tidal effects.

The amplitude of the variation provides insight to the star deformation: since the luminosity variation is around $0.3\%$, the star must be stretched by at least $0.3\%$, a value that is modulated by the orbit inclination. Since the binary is non-eclipsing, it is difficult to determine this inclination.\\


The 4-day period is linked to a slight asymmetry in the ellipsoids, produced for example by the heating of the two hemispheres of the stars which are in front of each other. We can assume that the rotation of the stars is synchronized with the orbital period, which is compatible with the rotation velocity obtained in Sect. 2.5.
As can be seen in Fig.~\ref{figure:6}, this period affects the odd and even minima of the curve, which is natural for the heating process, but also the odd and even maxima. 
This means that the asymmetry is also caused by a luminosity difference between the forward and backward hemispheres of the stars, which is usually related to the O'Connell effect (O'Connell 1951, Davidge \& Milone 1984). 

 However, the curve is also very similar to that of a chemically peculiar star with an irregular chemical composition at the surface (as for example HD~50773, see Luftinger et al., this issue). From what is known about HD~51106, we can probably exclude this hypothesis: the star is classified as an Am star, as confirmed by the analysis of its abundances using FEROS, SOPHIE, and FOCES spectra. Since Am stars are not known to have measurable magnetic fields, our star is unlikely to exhibit magnetic spots or noticeable chemical inhomogeneities at its surface. \\

\subsection{HD~50747}
For this star system, the situation is more exciting because we are clearly dealing with a pulsating variable star. Since it is in a triple system, and the physical parameters of two of the stars are badly determined, the basic questions are:
 \begin{enumerate}
\item which of the stars is a variable?
\item what type of variable is it?
 \end{enumerate}

 We could even investigate the possibility of there being two variable stars in the system. The variability is dominated by two frequencies, (F1 and F2). As we also detected two frequencies that are the sum and differences of F2 and F1, this indicates that F1 and F2 belong to the same star, and weakens the likehood that there are two variable stars in the system. 

The brightest component is also probably not a pulsator, because it clearly falls outside any known variability regions in the H­R diagram, but is instead situated outside the blue edge of the $\delta$-Scuti instability domain.

 We can clearly exclude the hypothesis that the star is a hot variable (e.g., SPB), because the luminosity indicates that the stars in the close orbit are cooler than the component in rapid rotation.
On the basis of the mass derived from the orbit analysis and the range of frequencies present in the power spectrum, the most likely hypothesis is a $\gamma$-­Doradus star. Among all the variable stars of this particular class, we know that a large proportion belong to binary systems (Mathias et al. 2004; Henry et al. 2007). 

In the case of HD~50747, the best candidate variable is the more massive of the two close components: this latest assumption is based on the probable mass of the star (Table~\ref{table:3}), somewhat heavier than the Sun,
whereas the other component seems to be lighter. Of course, if the value of sin $i$ is small enough, the lighter component could have a mass in the right range to be unstable. It is unfortunately difficult to determine  the fundamental parameters of the two close components accurately from the composite spectrum.  From examination of the spectra of HD~50747, it appears that this lighter component is very faint (its lines are scarcely visible), and it is unlikely to be hot enough to be in the instability strip.

 \section{Conclusion}
We have analysed observations of HD~51106 and HD~50747 acquired by CoRoT during its initial run.  
 To interpret these data, we also employed the analysis of pre-­CoRoT observations, obtained from the FEROS (ESO observatory), SOPHIE (Haute-Provence observatory), and FOCES (Calar Alto Astronomical Observatory) spectrographs: from this data we discovered that  HD~51106 is a binary system consisting of two Am stars, and HD~50747 is a triple system with two close components.

The light curve of HD~51106 shows no indication of intrinsic variation of the stars, because the observed quasi-sinusoidal variation can be attributed entirely to the orbital motion of the binary. 
HD~51106 is a very classical case of an Am star: most Am stars are in binary systems, the abundances anomalies being explained by the deceleration in the rotation caused by tidal effects. The usual view is that diffusion produces a settling of helium and prevents the excitation of pulsations (Vauclair et al., 1974).
Despite this phenomenon, several pulsating classical Am stars have been discovered (Kurtz 2000, Li 2000, Henry \& Fekel 2005). Even with its improved detection by CoRoT, HD~51106 remains, however, similar to the majority of non-pulsating Am stars.

Our analysis of the spectra of HD~50747 has demonstrated that it is a triple system, with two close components orbiting in 9.25 days and a third star dominating the luminosity.
 The light curve provides evidence of many frequencies between a few hours and one day. The most likely present interpretation is that one star in the system is a variable, the most probable candidate being the more massive of the two close components: its characteristics, although not determined precisely, and the range of frequencies observed are compatible with those of a $\gamma$-­Doradus star.

\begin{acknowledgements}
The FEROS data are being obtained as part of the ESO Large Program LP178.D-0361 (PI: E.Poretti). This work was supported by the italian ESS project, contract ASI/INAF I/015/07/0, WP03170.
The SOPHIE data are from observations at Observatoire de Haute Provence (CNRS), France.
KU acknowledges financial support from a European Community Marie Curie Intra European fellowship, number MEIF-CT-2006-024476.
PJA acknowledges financial support from the "Ram\'on y Cajal" postdoctoral fellowship programme of the Spanish Ministry of Education and Science.
ND and AH acknowledges V.Tsymbal for useful discussions and private communication. 
This work has made use of the ADS-CDS databases (CDS, Strasbourg, France), and of GAUDI, the data archive and access system of the 
ground-based asteroseismology programme of the COROT mission. The 
GAUDI system is maintained at LAEFF. LAEFF is part of the Space 
Science Division of INTA.

\end{acknowledgements}

\end{document}